\documentclass[%
 reprint,
 amsmath,amssymb,
 aps,
pre,
showkeys,
]{revtex4-2}
\usepackage{graphicx}
\usepackage{xcolor}
\begin{document}
\title{Probing the radiation-dominated regime of laser-plasma interaction\\
in multi-beam configurations of petawatt lasers}

\author{T.V. Liseykina}
    \email[Correspondence email address: ]{tatiana.liseykina@gmail.com}
    \affiliation{Institute of Computational Mathematics and Mathematical Geophysics SB RAS\\ 
Lavrentiev ave. 6, Novosibirsk, 630090, Russia}
    \affiliation{Institute of Applied Physics RAS Ul'yanov Street 46, 603950, Nizhny Novgorod, Russia}

\author{E.E. Peganov}
    \email[Correspondence email address: ]{egorpeganov.mephi@gmail.com}
    \affiliation{National Research Nuclear University MEPhI, Kashirskoe shosse 31, 115409, Moscow, Russia}

\author{S.V. Popruzhenko}
    \email[Correspondence email address: ]{sergey.popruzhenko@gmail.com}
    \affiliation{National Research Nuclear University MEPhI, Kashirskoe shosse 31, 115409, Moscow, Russia}
    \affiliation{Prokhorov General Physics Institute RAS, Vavilova 38, 119991, Moscow, Russia}
    \affiliation{Institute of Applied Physics RAS Ul'yanov Street 46, 603950, Nizhny Novgorod, Russia}

\date{\today} 

\begin{abstract}
We model numerically the ultrarelativistic dynamics of a dense plasma microtarget in a focus of several intersecting femtosecond laser pulses of multi-petawatt power each.
The aim is to examine perspective future experimental approaches to the search of the Inverse Faraday Effect induced by radiation friction.
We show that multi-beam configurations allow lowering the single beam peak laser power required to generate a detectable quasi-static longitudinal magnetic field excited due to the radiation reaction force.
The effect is significant at angles around $10^{\rm o}$ between the beam propagation axes, almost vanishes when the angle exceeds $20^{\rm o}$, and remains rather stable with respect to variations of relative phases and amplitudes of the beams. 
Quantum recoil accounted 
within semi-classical approach is shown to considerably suppress the longitudinal magnetic field, which however remains sizable.
We conclude that using four infrared femtosecond linearly polarized pulses, 15 petawatt power each, crossing at angles $\approx 10^{\rm o}$, the radiation-dominated regime of laser-plasma interaction can be experimentally demonstrated.
\end{abstract}


\maketitle

\section{Introduction}\label{sec:intro}

Effects of radiation friction or, equivalently, the radiation reaction force in dynamics of charged particles, beams and plasma subject to intense electromagnetic fields have remained in the scope of interest of strong-field physics for several latest decades.
Recently physics of the radiation reaction force (RRF) has become even more topical, thanks to the considerable progress in design and construction of multi-petawatt laser sources potentially capable to produce ultrashort electromagnetic pulses of intensity $10^{23}$W/cm$^2$ and higher (see reviews \cite{gonoskov_rmp22, fedotov_pr23, popruzhenko_usp23} for information on the current status of high-power laser facilities).
The theoretical framework for treatment of the RRF in classical electrodynamics has been amply developed \cite{Landau2, Jackson} in the past, and various signatures of radiation reaction in the dynamics and radiation of relativistic particles in external fields such as radiation damping in accelerators or radiation losses for cosmic rays had become tutorial examples. 
Recent experimental advances in the observation of interactions of ultrafast charged particles with crystals in the regime of channeling have opened the way to detect effects of quantum radiation reaction on the motion of positrons due to strong electric fields in aligned crystals \cite{dipiazza_ncomm18, dipiazza_prr19}; see also \cite{dipiazza_plb17, khokonov_plb19} for the theoretical treatment.
This research has  set an important milestone showing clear signatures of the quantum radiation reaction regime.
Together with the recent experiments on the interaction of laser-accelerated electron bunches with strong counter-propagating laser pulses \cite{cole_prx18, poder_prx18}, these studies have considerably extended knowledge of the realm of the quantum interaction regime, although its theoretical treatments still remain incomplete, and the transition from the fully quantum to the classical description is waiting for the ultimate clarification \cite{macchi_phys18, dipiazza_ncomm18}.

Despite the aforementioned recent experimental and theoretical advances, the subject remains topical and essentially intricate because of the two circumstances raising at the overlap of plasma physics and that of extreme electromagnetic fields.
Firstly, in a dense plasma driven by extremely strong electromagnetic fields of short focused laser pulses, the RRF is expected to significantly alter the collective dynamics providing conditions for a new interaction regime called \emph{radiation-dominated plasma dynamics} and enriched by new mechanisms of particle acceleration and quasi-static magnetic fields generation.
Secondly, applying intensities above $10^{24}-10^{25}$W/cm$^2$ or using pre-accelerated plasma interacting with pulses of lower intensity one may enter the multi-particle quantum regime of radiation reaction characterized by the development of quantum cascades and the creation of dense electron-positron plasmas, see e.g. \cite{gelfer_pra15, kostyukov_prl11}. 
Experiments in this intensity domain can provide fundamental knowledge useful for the development of a comprehensive quantum theory of radiation reaction.
For the detailed description of this research field we send the reader to reviews \cite{bulanov_rmp06, blackburn_rmpp20, gonoskov_rmp22, fedotov_pr23, popruzhenko_usp23} and the literature quoted therein.

The Inverse Faraday Effect (IFE) \cite{pitaevskii_jetp61, pershan_pr63, van_prl65} induced by the RRF in collisionless plasma interacting with circularly polarized (CP) electromagnetic radiation of extreme intensity can serve as an example of the aforementioned collective plasma phenomena arising due to radiation friction \cite{macchi_njp16,poprz_njp19,poprz_ejpp21}.
The IFE provides a new mechanism for the excitation of strong quasi-static magnetic fields in plasma, and, more importantly, it can be used as a benchmark of the radiation-dominated dynamics in classical or quantum regimes of laser-plasma interactions.
It was shown in Ref.\cite{macchi_njp16} that the synchrotron-like emission by fast electrons in a CP laser pulse results in a considerable irreversible angular momentum transfer from the laser field to the plasma followed by the excitation of ring-like electron currents in the polarization plane, which in turn generate a quasi-static magnetic field.
At laser intensities $\simeq 10^{24}$W/cm$^2$, this magnetic field can reach an amplitude of several gigagauss (GG).
More precisely, for the interaction of a CP pulse of $\lambda=800$nm wavelength and a super-Gaussian lateral intensity distribution of width $r_0=3.8\lambda$ the generation of a 1~GG magnetic field was predicted in the simulation \cite{macchi_njp16} for the peak intensity $J=4.4\cdot 10^{23}$W/cm$^2$.
The magnetic field exists for more than 100~fs in the volume $\simeq 50\lambda^3$ and should be accessible for detection by proton diagnostics.
 Further calculations made with a semi-classical account of the quantum recoil effect on the dynamics and radiation of plasma electrons \cite{poprz_ejpp21}, showed, for these interaction parameters, considerable suppression of the IFE without, however, qualitative changes in the picture.

The aforementioned interaction parameters can be reached at the peak laser power $\approx 80$PW and $\approx 1.2$kJ pulse energy.
However, despite the significant advances in petawatt (PW) laser technologies demonstrated in the two past decades, such values are still far from reach in laboratories.
The peak power of 10-15~PW at a pulse duration 20-30~fs looks as an optimistic promise for the nearest future.
The corresponding energies are 100-300~J per pulse.
Several laser facilities \cite{apollon,corels,ELI-BL,ELI-NP,SULF} are filling currently the interval of 1-10~PW power.
With this power level secured, projects of multibeam sub-exawatt laser installations will become realistic.
The laser facility XCELS combining up to 12 synchronized 10-15~PW laser beams can serve an example of such projects \cite{XCELS}, which can be put in realization within the nearest decade. 
Note that presently several possible multibeam configurations remain under discussion for the design of the future XCELS installation. 
Apart of the mentioned $12\times 15$PW scheme, there is another, which involves fewer beams 50PW each.

The possibility to superimpose coherently several laser beams in the focal area will allow lowering the single pulse energy and power thresholds required to achieve the radiation dominated regime, including the generation of strong magnetic fields through the IFE.
The muti-beam scheme would also eliminate the necessity of converting an initially linearly polarized (LP) beam into a CP one as it will be sufficient to use several pairs of LP beams with mutually orthogonal polarizations and an appropriate phase shifts propagating at small angles to each other.
In this work, we examine such a scheme numerically by modeling the laser-plasma interaction within the particle-in-cell (PIC) method using the codes UMKA \cite{umka,umka_1998,NIC_2020} and SMILEI \cite{smilei}.
The former allows for simulating the plasma dynamics both in the classical and quantum regimes of interaction with the effect of photon recoil accounted for semi-classically in the second case along the method of Refs.\cite{ritus_jslr85,kirk_ppcf09}.
We aim at the following objectives: (a) to estimate the minimal single pulse energy and power required for the generation of a significant, detectable magnetic field due to the radiation reaction effect in a multi-beam configuration and (b) to check the sensibility of the generated magnetic field peak amplitude and spatial distribution to the relative phase shifts of the beams, variations in their amplitudes and to the angle at which the beams intersect.
The robustness of the scheme with respect to the first two factors is of particular importance, since they are difficult to control in multi-PW sources with a low repetition rate.
The angle between the beams' propagation directions needs optimization because the electromagnetic field structure close to CP vanishes quickly with this angle increasing, while using nearly parallel beams intersecting at small angles is technically difficult. 
The output of our study will help to formulate requirements to the architecture of multi-beam multi-PW laser source XCELS and other laser facilities of similar class.

The structure of the paper is following.
In the next section we formulate the statement of the problem and discuss parameters of the laser pulses and plasma.
Section \ref{sec:Results} is dedicated to presentation and discussion of numerical results.
Section \ref{sec:Conclusion} reports conclusions and outlook.
Preliminary results of our calculations related to the two-beam scheme were reported in \cite{liseykina_bull23}.

\section{Statement of the problem\label{sec:Statement}}
\subsection{Interaction scheme and parameters}

\begin{figure*}[ht!]
\includegraphics[width=6.8in]{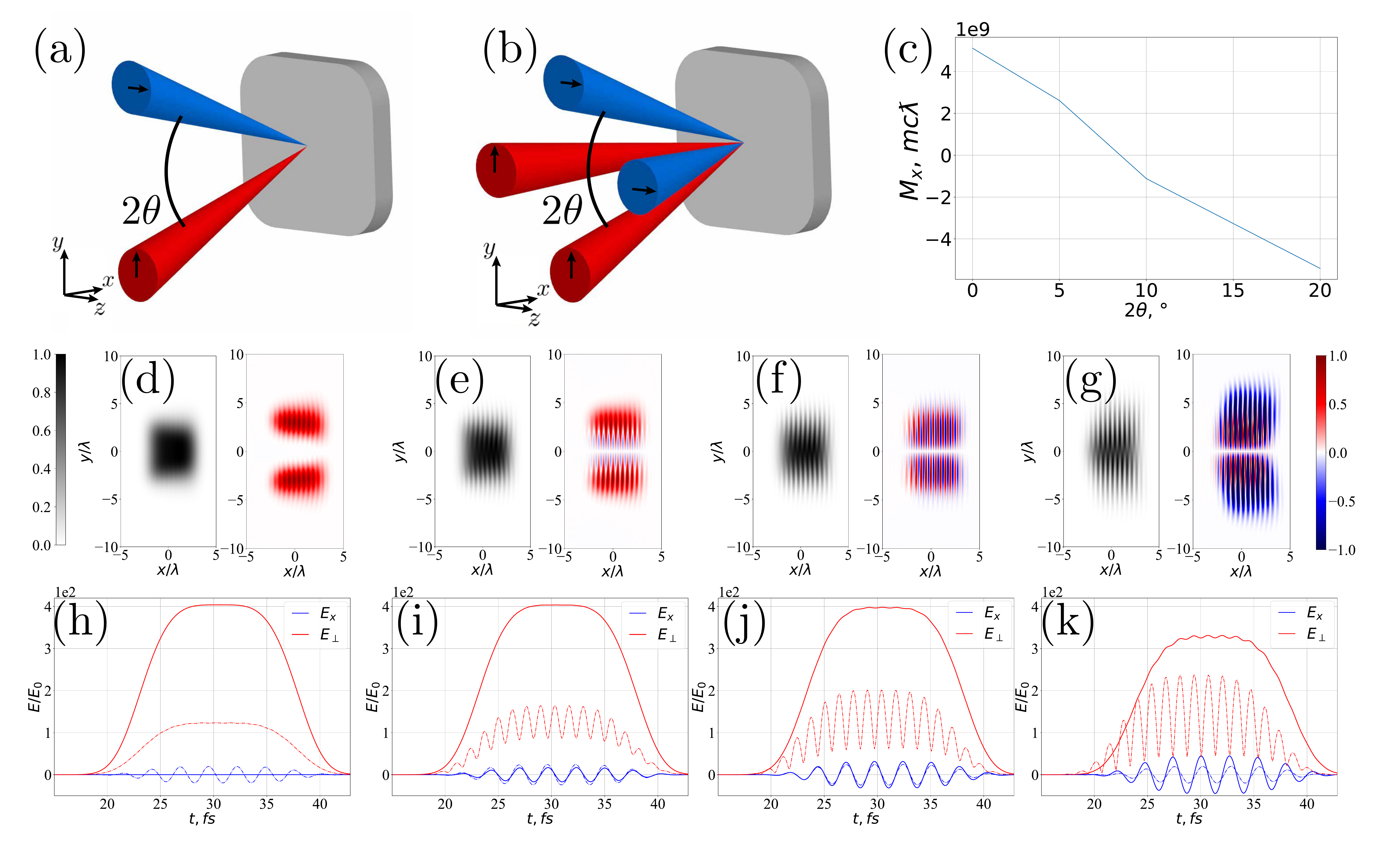}
\caption{First row: sketch of the 2-beam (a) and 4-beam (b) setups; (c) projection of the total angular momentum $M_x$ of the laser field on the axis $x$ perpendicular to the plasma slab as function of $\theta$ for the 4-beam case. The angular momentum is measured in units of $mc\lambdabar\approx 2.5\cdot 10^{22} {\rm erg}\cdot {\rm s}$.
Second row: instant distributions in $a^2(x,y)$ (left panels, gray-scale) and $m_x$ (right panels, color-scale) for $2\theta=0^{\rm o}$ (d), $2\theta=5^{\rm o}$ (e), $2\theta=10^{\rm o}$ (f) and $2\theta=20^{\rm o}$ (g) for the 4-beam case. Each distribution is normalized to the maximum value of the plotted quantity.
Third row: time dependence of $E_x/E_0$ (blue lines) and $E_\perp=\sqrt{E_y^2+E_z^2}/E_0$ (red lines) at points (0, 0, 0) (solid lines) and (0, \(4\lambda\), 0) (dash-dotted lines) for the same set of angles as in the second row.
}
\label{fig:Fig1}
\end{figure*}
We consider the interaction of two or four LP laser beams of $\lambda=800$nm wavelength with a helium fully pre-ionized plasma slab of a $10\lambda$ thickness and initial concentration $n_0=1.55\cdot 10^{23}$cm$^{-3}=90n_c$ with 
\begin{equation}
    n_c=\frac{m_e\omega^2}{4\pi e^2}
\end{equation}
being the critical concentration for the laser frequency $\displaystyle\omega=2\pi c/\lambda$; $m_e$ and $e$ are the electron mass and elementary charge correspondingly.
We characterize the electromagnetic field amplitude $E_0$ by its invariant dimensionless form
\begin{equation}
    a_0=\frac{eE_0}{m_ec\omega}
    \label{a0}
\end{equation}
with $c$ being the speed of light. 
A position-dependent analogue $a(\bf r)$ of (\ref{a0}) is defined by replacing the field amplitude $E_0$ with $E(\mathbf{r})=\sqrt{<E^2>(\mathbf{r})}$, where $<...>$ denotes time averaging.
The plasma parameters are chosen to keep them close to a set of earlier calculations \cite{macchi_njp16, poprz_njp19, poprz_ejpp21}, where the IFE in a single CP beam of intensity $J\approx 10^{24}$W/cm$^2$ and higher was studied.
For LP pulses of frequency $\omega =2.4\cdot 10^{15}$s$^{-1}$, $a_0=1$ at $J=2.1\cdot 10^{18}$W/cm$^2$.
In a single CP laser beam, our previous calculations \cite{macchi_njp16, poprz_njp19} did not reveal any significant longitudinal magnetic field for $a_0<200$.
Thus in the present calculation we investigate the case of intensities $\approx 8\cdot 10^{22}$W/cm$^2$ in a single beam.

The laser pulse is initialized by setting a non-stationary boundary condition on the left boundary $x=x_0=-5\lambda$ of the computation domain.
The electromagnetic field inside this domain is found by solving the Maxwell's equations numerically. 
To set the boundary condition, we use super-Gaussian beams 
of the following shape
\begin{equation}
     f(x',y',z',t)= \exp\left\{-\frac{(y'^2+z'^2)^2}{r_0^4} -\frac{(x'-ct)^4}{r_c^4}\right\}
\label{fields_b}
\end{equation}
The finite-difference time-domain method \cite{Yee,Langdon1976} with the total-field/scattered field formulation \cite{Taflove2005} in x-direction is used to propagate the electromagnetic fields.
Coordinates $(x',y',z')$ in (\ref{fields_b}) are introduced for each beam individually such that the wave vectors ${\bf k}'=~(k,0,0)$ are orthogonal to the corresponding $(y',z')$ planes.
Primed coordinates are related to laboratory ones used in the simulation by rotation in an angle \(\theta\) in the $(x, y)$ plane for the two-beam setup and in the $(x, y)$ and $(x, z)$ planes for the four-beam setup.  
Schematic representation of both setups is shown on Fig.~\ref{fig:Fig1}(a,b).
The beam lateral waists and the longitudinal pulse lengths are taken $r_0 = 3.8\lambda$ and $r_c = 3\lambda$ correspondingly.
The effective pulse duration is $\tau=3\lambda/c=6\pi/\omega\approx 7.2$~fs.
This is roughly three times shorter than a typical pulse duration for most of the multi-PW sources. 
This difference is not of crucial importance for our analysis, while the application of such short pulses considerably accelerates demanding numerical simulations.

The boundary condition \eqref{fields_b} does not focus the field on the plasma surface, instead the beams appear at $x=0$ defocused compared to the plane $x=-5\lambda$. 
This defocusing is relatively weak thanks to a small distance between the boundary and the plasma surface.
Thus the peak intensity estimated using of the field distribution (\ref{fields_b}) on the boundary does not differ much from the value reached on the plasma slab surface.
Solutions found in free space (without plasma) allow to check, how quickly the field in the area where the beams overlap deviate from CP with the angle $\theta$ increasing.
We characterize this deviation by the distributions in dimensionless intensity ${\tilde J}=a_0^2({\bf r})$ and in the density of x-projection of angular momentum ${\bf m}({\bf r})=[{\bf r}[{\bf E}{\bf B}]]/4\pi c$ shown in Fig.~\ref{fig:Fig1}(d-g).
For a CP beam the $a^2$-distribution is smooth (Fig.~\ref{fig:Fig1}(d)), and the $m_x$ density is everywhere positive, while for $\theta\approx 10^{\rm o}$ a considerable difference from the CP case is already clearly seen.
The sign of the integrated projection  of the angular momentum of the field $M_x$ flips at $2\theta\approx 8^{\rm o}$ (Fig.~\ref{fig:Fig1}(c)).
The time dependence of the electric field projection $E_x$ and of the absolute value of its lateral component $E_{\bot}=\sqrt{E_y^2+E_z^2}$ shown in panels of Fig.~\ref{fig:Fig1}(h-k) also demonstrates the deviation from CP increasing with the growth of the convergence angle. 
For $2\theta>10^{\rm o}$ the field structure remains close to CP on the axis $y=z=0$, while at $y=4\lambda$ the deviation is apparent.
Taking into account that the IFE is induced by electron currents excited near the laser beam lateral edge ($\approx 3-4\lambda$ in our case) \cite{macchi_njp16} this deviation should considerably suppress the magnetic field generation.
Our further simulations (not present in the plot) show that the field structure in the overlap area of the beams depends on the boundary condition.
In particular, a Gaussian boundary condition defined such that the beams overlap in their focal waists leads to a slower decrease in the degree of CP with $\theta$ increasing.
However, this does not change our results qualitatively. 
Summarizing the information delivered in Fig.~\ref{fig:Fig1}, we conclude that with $\theta$ increasing, the field structure in the focus deviates form CP rather quickly so that the case $\theta>10^{\rm o}$ is not of interest for the detection of IFE signatures \cite{liseykina_bull23}.

Numerical runs described in the next subsection were arranged such that the laser pulse front edge arrives to the front surface of the plasma slab $x=0$ at $t=0$ so that $t>0$ is time past after the start of the interaction. 
With the efficient pulse duration $\tau=3T_L\approx 7.2$fs the laser field is being entirely reflected by the plasma during $\tau_{\rm int}\approx 15-20$fs, which can be considered as the full interaction time.
We look at the magnetic field excited in the plasma after the interaction,  typically at $t=30-40T_L$.
For the super-Gaussian pulse (\ref{fields_b}) the peak power is
\begin{equation}
P=\frac{\pi^{3/2}}{2\sqrt{2}}J_0a_0^2r_0^2
\label{P}
\end{equation}
with $J_0=2.1\cdot 10^{18}{\rm W/cm}^2$ for $\lambda=800$nm and LP.
Taking $a_0=200$ one obtains from (\ref{P}) $P\approx 16$PW, the value close to that expected at the XCELS laser facility under development \cite{XCELS}.
The full pulse energy is about 150J.
Note that (\ref{P}) only gives an estimate of the peak laser power, because the field structure in the interaction volume obtained through the numerical solution of the Maxwell equations deviates from the analytic form (\ref{fields_b}).

\subsection{Numerical methods\label{Method}}

The interaction scenario described above
was numerically modeled using two independently developed PIC codes  UMKA \cite{umka,umka_1998,NIC_2020} and SMILEI \cite{smilei}.
The two codes provide similar though not identical results for the number of nodes not less than 32 per laser wavelength and a sufficiently large number of macroparticles per cell.

In the simulations, 64 macroions and macroelectrons per cell were used at the spatial resolution $\Delta x=\Delta y=\Delta z=~\lambda/40.$ 
Boundary conditions for the fields on the left plane of the computational domain, $x=-5\lambda$, are given by (\ref{fields_b}). Otherwise, the conditions for the fields and particles are: absorption in the propagation direction and periodicity in the lateral directions. 
The size of the simulation box $[35\times 25\times 25]\lambda^3$ is chosen sufficiently large to rule out any boundary effects
on the particles' and fields' distributions in the central part of the simulation domain. 

In both codes, the radiation reaction force  $\mathbf{F}_{RR}$ is included in the equations of motion for electrons in the form
\cite{Landau2} 
\begin{equation}
\begin{split}
\mathbf{F}_{RR} = -\frac{2}{3}r_e^2&\left[ \gamma^2\left(\mathbf{F}_L^2-\left(\frac{\mathbf{v}}{c}\cdot\mathbf{E}\right)^2\right)\frac{\mathbf{v}}{c}\right.\\
&\left.-\mathbf{F}_L\times\mathbf{B}-\left(\frac{\mathbf{v}}{c}\cdot\mathbf{E}\right)\mathbf{E}\right]~.
\label{LL}
\end{split}
\end{equation}
Here, $r_e= e^2/mc^2\approx 2.8\cdot 10^{-13}$cm is the classical electron radius,  $\mathbf{v}$ is the electron velocity, $\displaystyle \mathbf{F}_L=\left(\mathbf{E}+\mathbf{v}/c\times\mathbf{B}\right)$ 
and $\gamma=1/\sqrt{1-v^2/c^2}$ is the relativistic gamma-factor. 
This is a reduced expression for the RRF, compared to its full Landau-Lifshitz form \cite{Landau2}.
It follows from the latter after omitting terms (small in the considered case) containing time and space derivatives of the electromagnetic fields.

Eq.(\ref{LL}) treats radiation reaction classically.
At the considered values of the dimensionless field amplitude $a_0=200-400$, classical description may appear insufficient, depending on the details of interaction including its geometry.
Analysis of quantum effects in photon emission in intense electromagnetic fields and and their contribution to the RRF can be found in reviews \cite{blackburn_rmpp20, fedotov_pr23} and papers \cite{ritus_jslr85, kirk_ppcf09, blackburn_pra24}.
In our case, light pressure and the ponderomotove forces induced by the strong laser pulse result in a considerable longitudinal plasma acceleration, leading to predominantly co-propagative interaction geometry.
This in turn suppresses the effective value of the quantum parameter $\chi=E^{\prime}/E_{\rm cr}$, which is the ratio of the field amplitude in the electron's rest frame to the critical field of quantum electrodynamics $E_{\rm cr}=m^2c^3/e\hbar=1.32\cdot 10^{16}$V/cm.
As a result, the interaction proceeds in the  semi-classical regime, $0.1<\chi<1$, where quantum recoil may play a significant role, but typically does not qualitatively alter results.
The impact of quantum recoil on the IFE driven by the RRF was considered in our earlier publication \cite{poprz_ejpp21}.
Following the approach explored there, we apply a semi-classical modification of the RRF force via the $g$-factor introduced by Ritus \cite{ritus_jslr85, ritus}, with the following interpolation \cite{thomas-prx12}
\begin{equation}
g(\chi)=\bigg(1+12\chi+31\chi^2+3.7\chi^3\bigg)^{-4/9}.
\label{g-chi}
\end{equation}
The quantum parameter $\chi$ is calculated at each time step for each macroelectron by taking the values of the electromagnetic 
field at the macroelectron position. 
The quantum-corrected RRF reads $\tilde{\mathbf{F}}_{RR}=g\mathbf{F}_{RR},$ with $\mathbf{F}_{RR}$ given by Eq.\eqref{LL}.

The codes use a numerical algorithm  based on the assumption that the acceleration of particles in their rest frame is dominated by the Lorentz force (see Ref. \cite{Tamburini_2010} for detail). 
The numerical implementation allows adding radiation reaction effects to any PIC code based on the standard Boris pusher algorithm for the calculation of the particles acceleration with small computational costs.
Inclusion of dissipative radiation losses  through the RRF in a PIC code is justified if the following conditions are met: 1) characteristic frequencies of synchrotron-like radiation emitted by the plasma electrons are much higher than the highest frequency that can be resolved on the numerical grid; 2) radiation at such frequencies is incoherent; 3) the plasma is transparent to electromagnetic fields of these frequencies.
For the present calculations, these requirements are well satisfied (see estimates in Refs.\cite{macchi_njp16, poprz_ejpp21}).
At lower though still ultra-relativistic intensities, radiation responsible for the RRF may become partially coherent, which requires to revisit the description of radiation reaction effects.
For this specific case, see recent publications \cite{gelfer_mre24,gelfer_prr24}. 

\section{Results and discussion\label{sec:Results}}
\begin{figure*}[t!]
\includegraphics[width=6.8in]{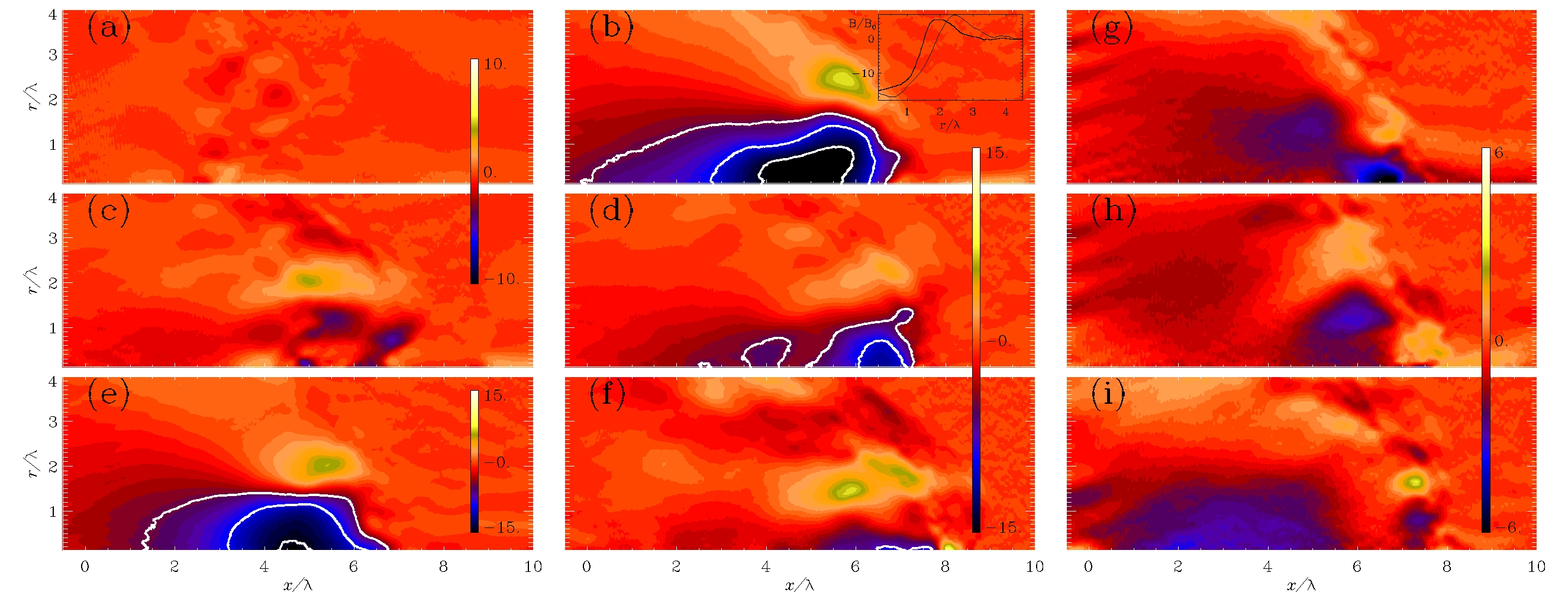}
\caption{Spatial distributions in $B_x/B_0$ as functions of $x$ and $r=\sqrt{y^2+z^2}$ at $t=35T_L$ for the cases of two (a,c) and four (b,d,f-i) beams with the angle of convergence (see Fig.~\ref{fig:Fig1} and text around for explanations) $2\theta=5^{\rm o}$ (a), (b), (g); $2\theta=10^{\rm o}$ (c), (d), (h) and $2\theta=20^{\rm o}$ (f), (i). The single beam power was fixed by 16PW corresponding to $a_0=200$. For reference, panel (e) shows the distribution for a single CP beam with power $P=64$PW equal to the total power of the four beams. Panel (c) gives the distribution for the same case as (a), but assuming 32PW $(a_0=280)$ in each beam instead of 16~PW. Note different color scales in panels (a,c), (b,d,e,f) and (g-i) correspondingly. The lines of constant magnetic field $B_x=[-15,-10,-5]B_0$ are shown white.  The inserts in frames (b,g-i) display the one-dimensional radial profiles of $B_x$ for the distributions shown in frames (b) at $x=5.3\lambda$ (thin line) and in frame (e) at $x=5\lambda$ (thick line).
 Panels (g,h,i) show the counterparts of distributions (e,b,d) calculated with the quantum $g$-factor included. }
\label{fig:Fig2}
\end{figure*}
Our discussion is centered around magnetic field distributions in space for (a) different number of the beams, (b) different angles of convergence $\theta$ and (c) the case of fluctuations in relative phases and amplitudes of the beams.
All distributions shown in plots correspond to the time instant $t=35T_L\approx 85$fs after the beginning of the interaction, i.e. more than 50fs after the laser pulse is reflected.

Firstly, we examine the efficiency of the magnetic field excitation in different geometries.
To this end, we compare the distributions of the longitudinal magnetic field $B_x$ recorded at $t=35T_L$ for the 2- and 4-beam cases at different convergence angles.
The magnetic field amplitude is normalized to $B_0=m_ec\omega/e=1.38\cdot 10^8$Gauss.
To reduce fluctuations and make the signal smoother, we integrate the distributions over angle in the $(y,z)$ plane so that the field $B_x$ in Fig.~\ref{fig:Fig2} depends on the longitudinal coordinate $x$ and the radial one $r=\sqrt{y^2+z^2}$.

From these distributions, we gain that:
\begin{enumerate}
    \item When the IFE is present for collinear beams, it survives for the angles of convergence up to $2\theta \simeq 10^{\rm o}$. For $2\theta \simeq 20^{\rm o}$ the magnetic field drops down considerably, both in the amplitude and the occupied space.
These features are similar for two and four beams setup. For the two beam scheme a similar behavior was reported in a brief publication \cite{liseykina_bull23}.
    \item The difference between the 2- and 4-beam schemes is evident for the fixed power of a single beam. The distributions shown in Fig.~\ref{fig:Fig2}(a,c) demonstrate that for the 2-beam scheme the threshold power, when some signatures of the magnetic field generated through the IFE can be expected, is $\simeq 30$PW, well above the presently achieved level of 5PW and the soon expected that of 15PW. Instead, the 4-beam scheme should work reliably at the 15PW power. For the scale $B_x\in [+10,-10]B_0$ of panels (a) and (c) no magnetic field is visible when two 15PW beams are applied. Reduction of the interval reveals some fields $B_x$ which, however, lie beyond the reliable signal level of the present PIC simulations.
    \item The quasi-static longitudinal magnetic field $B_x$ vanishes in correlation with the decrease in the longitudinal projection of the field angular momentum (see Fig.~\ref{fig:Fig1}), which agrees well with macroscopic picture of the IFE \cite{macchi_njp16} based on the irreversible transfer of angular momentum from the field to the particles. However, one can notice from the comparison of Fig.~\ref{fig:Fig1}(c) and Fig.~\ref{fig:Fig2}~(d,f) that the magnetic field of negative sign survives even when the sign of $M_x$ is changed. An explanation of this extended survival of the IFE with the angle between the beams increasing stems from the distributions shown in Fig.~\ref{fig:Fig1}(f,g) and (h-k). It is clear from these plots that the density of $m_x$ remains positive in the central part of the interaction volume, up to $r\approx 3\lambda$, while the electron current generating the field is concentrated near $r\approx 2\lambda$. The lateral electric field, although oscillating at $y=4\lambda$, still contains a significant fraction of a CP component. The periphery of the interaction volume, which contributes a lot into the net angular momentum $M_x,$ plays a minor role in the excitation of the electron current. 
    \item
   At small values of $\theta$, the magnetic field can be even more efficiently excited than in the fully collinear geometry.
This counter-intuitive behaviour of the IFE can be noticed by comparing the magnetic field distributions shown in panels (b) and (e).
In the latter, we show the distribution for a single CP beam with $a_0=400.$ 
In this case, the maximal value $B_{x{\rm max}}=15.6B_0$ (thick line in the insert of Fig.~\ref{fig:Fig2}b), while for the distribution generated by four LP beams with $a_0=200$ each and $2\theta=5^{\rm o}$ $B_{x{\rm max}}=17.3 B_0$ (thin line in the inset).  
A more significant difference between the two cases is that the magnetic field occupies a considerably larger volume in the four beams setup. 
In Fig.~\ref{fig:Fig2}(b) the lateral size of the space where the magnetic field exceeds $15 B_0$ is more than three times bigger than that in Fig.~\ref{fig:Fig2}(e). This effect can be explained by examining the laser field structure on
the shoulders of the focus. 
The panels in the third row of Fig.~\ref{fig:Fig1} show the electric
field time evolution at the focus center (solid lines) and at a lateral distance of $4\lambda$, where the angular momentum density is the highest (see plots in the
second row of Fig.~\ref{fig:Fig1}). 
Comparison of panels Fig.~\ref{fig:Fig1}(h) and Fig.~\ref{fig:Fig1}(i) shows that while the
electric field at the focus center (solid lines) is almost the same for the cases of a single CP and of four LP pulses, the field at the focus periphery (dash-dotted lines) appears at its interference maxima higher for the four beam case. 
This interference effect, efficient enough taking into account that the magnetic field amplitude scales approximately as $a_0^4$ \cite{macchi_njp16}, makes the IFE stronger at larger distances from the beam axis.
At the same time, although the electric field amplitude at the periphery keeps growing with the angle increasing (see dash-dotted lines in panels Fig.~\ref{fig:Fig1}(j) and Fig.~\ref{fig:Fig1}(k)), the angular momentum density drops down there. 
Therefore, though $a_0$ is sufficiently large out of the axis for bigger angles, the field polarization state is far from circular, which suppresses the angular momentum transfer at large distances. 
As a result, for larger angles the part of the focus where the IFE remains efficiently
induced, shrinks down to the small vicinity of the x-axis. 

\item
In panels (g-i) of Fig.~\ref{fig:Fig2} we show results obtained for the same beam parameters as used for (e,b,d) but with the quantum recoil effect semi-classically accounted for through the factor $g(\chi)$ (\ref{g-chi}).
Quantum corrections considerably damp the radiation reaction force leading to a visible decrease both in the magnetic field amplitude and in the size of space where the magnetic field is observed.
The reduction factor is more than two (see also Fig.~2 in \cite{poprz_ejpp21}). 
However, the spatial structure of the generated field and its dependence on the angle between the beams remain similar to that 
observed in the classical simulations. 
Note that the account of recoil through the semiclassical $g$-factor may lead to an overestimation of quantum effects compared to the case in which they are treated stochastically \cite{Niel_2018,Niel_2018a} so that the magnetic field amplitudes in the counterpart panels (e)-(g), (b)-(h) and (d)-(i) should be considered as the upper and the lower bounds to the field, which could potentially be predicted in a full quantum calculation with stochastic radiation events and other quantum effects incorporated.
The field distribution in Fig.~\ref{fig:Fig2}(i) looks qualitatively different from that shown in all other panes.
Excitation of a relatively strong field at $2\lambda\le x\le 5\lambda$ can be attributed to a surface mechanism not connected with the RRF \cite{pukhov_njp21}.
This mechanism becomes more efficient with the angle between the pulse propagation and the vector normal to the surface increasing.
Discussed contribution is present in panel (d) too, in the same space region, but there it remains much less visible because of a stronger contribution from the IFE.

\end{enumerate}

\begin{figure*}[ht!]
\includegraphics[width=6.8in]{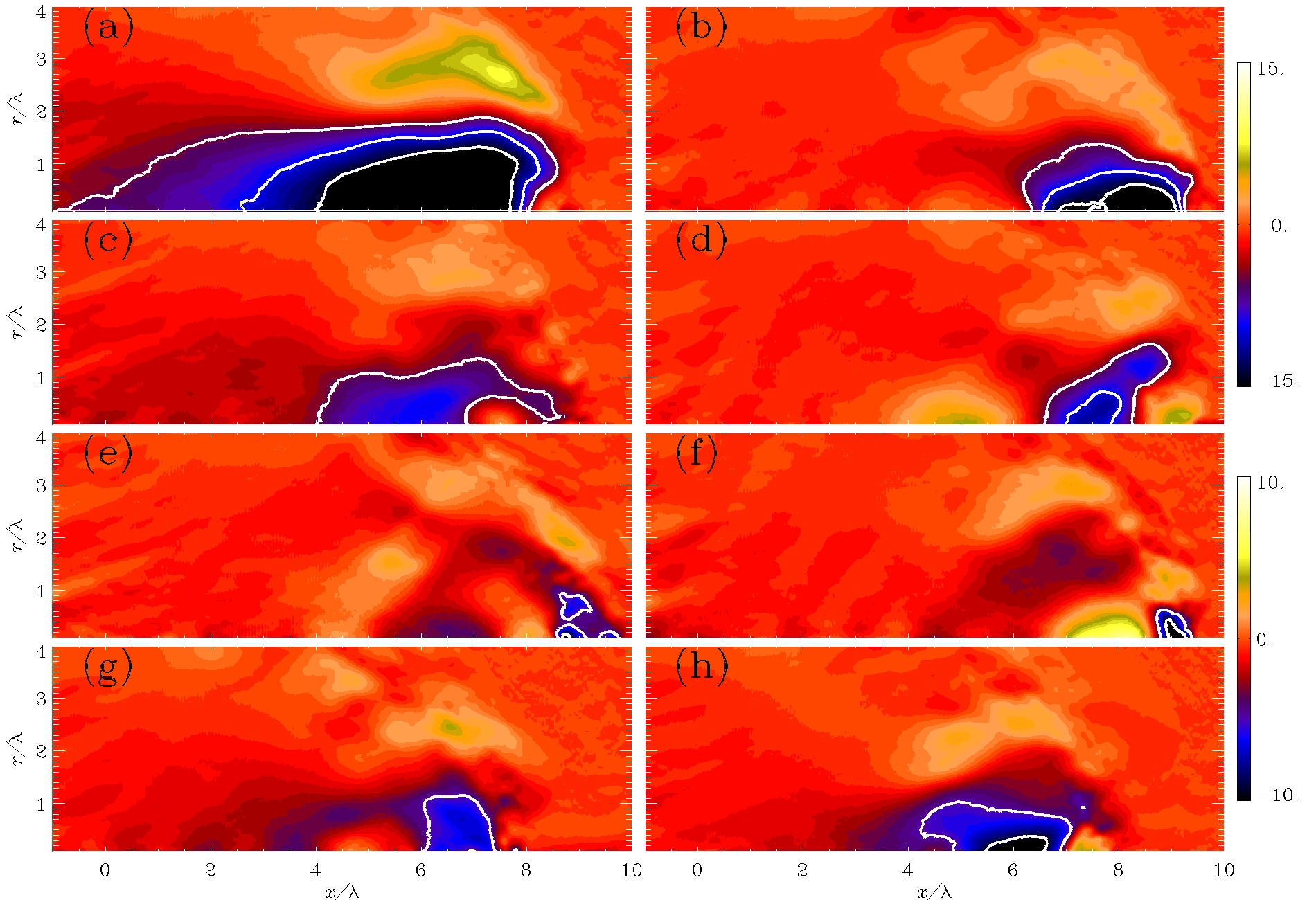}
\caption{Magnetic field distribution for the 4-pulse scheme with $2\theta=5^{\rm o}$ (a,c,e) and $2\theta=10^{\rm o}$ (b,d,f) calculated at $t=35T_L$ for the additional to $\pi/2$ phase shift $\Delta\varphi =0.0$ (a,b), $\Delta\varphi =0.2\pi$ (c,d) and $\Delta\varphi =0.4\pi$ (e,f). The field amplitude $a_0=250$ in each beam corresponds to the peak intensity $J\approx 1.4\cdot 10^{23}{\rm W}/{\rm cm}^2$ and power $P \approx 20$PW per beam. 
Panel (g) shows the distribution resulting from random relative phase shifts $0.075\pi, -0.018\pi, 0.05\pi$; panel (h) -- the distribution for randomly chosen amplitudes $a_0=184, 242, 162$ and $239$. The lines of constant magnetic field $B_x=[-15,-10,-5]B_0$ are displayed  white.}
\label{fig:Fig3}
\end{figure*}

Secondly, for multi-PW laser sources with sub-kJ pulse energies and low repetition rates, it is hard to expect the beams to be well synchronized in phase and amplitude.
For such laser systems under development, the option of fully incoherent beams is also considered as one of realistic scenarios.
Thus it is of high relevance for the future experiments to make a step out of the idealistic assumption of fully coherent in-phase beams and to study sensibility of the effect to fluctuations of the relative phases and amplitudes of the beams.
Fig.~\ref{fig:Fig3} presents the magnetic field distribution $B_x(x,r)/B_0$ recorded at $t=35T_L$ for the 4-beam case with $2\theta = 5^{\rm o}$ and $2\theta = 10^{\rm o}$. 
These cases were chosen basing on the analysis of the distributions present in Fig.~\ref{fig:Fig2}.
Panels (a-f) correspond to regular phase shifts between the pairs of beams with orthogonal polarizations, i.e. between the two blue and the two red beams in Fig.\ref{fig:Fig1}(b), while panels (g) and (h) show the distributions generated by the beams with randomly shifted phases and amplitudes, correspondingly.  

\begin{figure}[ht!]
\includegraphics[width=3.5in]{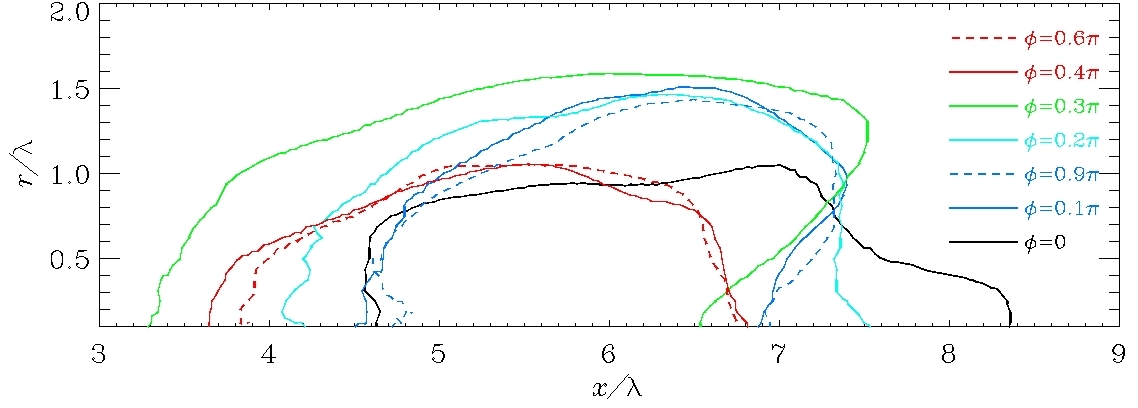}
\caption{Lines of constant magnetic field $B_x=-15B_0$ (for $\Delta\varphi<\pi/2$) and $B_x=15B_0$ (for $\Delta\varphi>\pi/2$) calculated for the 2-pulse scheme at $a_0=500$ and $2\theta=5^o.$}
\label{fig:Fig4}
\end{figure}

As expected, the efficiency of the magnetic field generation drops down with the phase shift increasing. 
This suppression can be qualitatively demonstrated by plotting contours of the constant magnetic field for different phase shifts.
The result is shown in Fig.~\ref{fig:Fig4}.
It is important to note that the effect survives in a broad interval of relative phases.
Comparison of panels (c), (d) (e) and (f) in Fig.~\ref{fig:Fig3} shows that the magnetic field remains visible at relative phases filling more than 50\% of the interval $(0,\pi/2)$.
The distributions in Figs.3 and 4 were calculated for higher field amplitudes $a_0=250$ and $a_0=500$.
In the latter case, the 2-beam scheme was considered.
A comparison of results presented in these Figures shows, in particular, that both the schemes are, on the qualitative level, equally sensitive to fluctuations of relative phases of the beams.
In experiment, both the relative phases and amplitudes of the beams will be difficult to control.
Panels (g) and (h) of Fig.~\ref{fig:Fig3} show the distributions for randomly chosen phase shifts and amplitudes.
These results also suggest that the scheme is acceptably stable with respect to these parametric uncertainties.

Thirdly, the degree of focusing also should play a certain role in the magnetic field excitation at the fixed pulse energy and power.
We examined the sensitivity of the scheme to the tightness of the laser focus. 
The one used in Figs.2-4 with $r_0=3.8\lambda$ can be reduced up to $r_0=2\lambda$ keeping it still realistic for the state-of-the art focusing systems applied in PW lasers, see, e.g. \cite{corels}.
Our simulations made with such tightly focused pulses showed a higher magnetic field or, equivalently, a lower power threshold per single beam.
However, for the tighter focusing and non-vanishing angles of convergence $\theta$, another mechanism of the magnetic field excitation starts playing an important role.
The contribution of this mechanism is clearly visible in Fig.\ref{fig:Fig2}(i) as a magnetic field located at $2\lambda \le x\le 5\lambda$.
It stems not from the effect of radiation friction, but from that of a sharp plasma edge, which leads to additional angular momentum transfer from the laser field to the plasma electrons at the laser incidence not normal to the plasma surface \cite{pukhov_njp21}.
In the forthcoming publication we will examine this mechanism and its relative significance with respect to the angular momentum transfer induced by radiation friction.

\section{Conclusion\label{sec:Conclusion}}

In conclusion, we have studied the possibility to reach the radiation-dominated regime of the laser plasma interaction and, in particular, to observe the Inverse Faraday Effect driven by radiation reaction in a multi-beam scheme.
Our numerical results show that application of four beams partially synchronized in phase and amplitude lowers the single beam power required for reliable detection of the quasi-static longitudinal magnetic field excited through the IFE to 15~PW, which looks a realistic value for several forthcoming laser facilities. 
Most importantly, the effect is shown rather robust with respect to fluctuations of relative phases and amplitudes.
This enhances the possibility of the observation at low repetition rates typical for multi-PW laser sources. 
The effect of quantum recoil of emitted photons reduces the radiation reaction force and results in a significant suppression of the excited quasi-static magnetic field, which however remains clearly visible for all considered sets of parameters of the 4-beam scheme.

An important feature observed in our simulations is the relatively low decrease of the effect with angle between the beams increasing.
The latter leads to the following changes in the electromagnetic field in the focal area: (i) the total angular momentum carried by the field decreases (see Fig.~\ref{fig:Fig1}(c)) -- this is a global effect, and (ii) the spatial structure of the field deviates from that of a CP electromagnetic wave -- a local effect. 
Our simulations show that the local field structure appears more important for the magnetic field excitation than the integrated angular momentum value. This can be clearly seen from comparison of e.g. Fig.~\ref{fig:Fig1}(e) and Fig.~\ref{fig:Fig1}(g) for $2\theta=5^o$ and $2\theta=20^o,$ correspondingly. The value of $M_x$ changes the sign between these two points and
even more, the absolute value of (negative) $M_x$ is higher for $2\theta=20^o$ than for $2\theta=5^o.$
However, some magnetic field of the same sign is still present in Fig.~\ref{fig:Fig2}(f) and in Fig.~\ref{fig:Fig2}(i). 
It is tightly localized near the axis $r=0,$ which agrees well with the fact that the angular momentum density remains positive in Fig.~\ref{fig:Fig1}(g) close to the axis, too. 
Thus the IFE induced by radiation friction is rather sensitive to the angular momentum density than to its integrated value. This allows
us to estimate the effective volume where a strong magnetic field can be created.

At the future XCELS facility \cite{XCELS}, installations with up to 12 beams are considered.
On one side, increasing the number of beams will obviously lower further the power threshold required for IFE observation.
On the other side, taking into account the constraints on the angle $\theta$ and effects of the relative phase shifts we examined in this paper, application of $\sim 10$ or more beams may appear highly problematic from the technical viewpoint. 
We conclude that the four beam scheme at the 10--15PW level is optimal for the search of the giant magnetic field excitation in the radiation-dominated regime.

\section*{Acknowledgment}
We are thankful to E. Gelfer and A. Fedotov for fruitful discussions and to the anonymous referee who provided useful and detailed comments on the manuscript.
Numerical simulations were performed at the Joint Supercomputer Center of the Russian Academy of Sciences, at the  Siberian Supercomputer Center of the Siberian Branch of the Russian Academy of Sciences, and at the Computing Center at the MEPhI Laboratory for Extreme Light Physics.
TVL and SVP acknowledge support of the Russian Science Foundation (grant No. 20-12-00077).


\bibliography{Literature}

\begin{thebibliography}{47}%
\makeatletter
\providecommand \@ifxundefined [1]{%
 \@ifx{#1\undefined}
}%
\providecommand \@ifnum [1]{%
 \ifnum #1\expandafter \@firstoftwo
 \else \expandafter \@secondoftwo
 \fi
}%
\providecommand \@ifx [1]{%
 \ifx #1\expandafter \@firstoftwo
 \else \expandafter \@secondoftwo
 \fi
}%
\providecommand \natexlab [1]{#1}%
\providecommand \enquote  [1]{``#1''}%
\providecommand \bibnamefont  [1]{#1}%
\providecommand \bibfnamefont [1]{#1}%
\providecommand \citenamefont [1]{#1}%
\providecommand \href@noop [0]{\@secondoftwo}%
\providecommand \href [0]{\begingroup \@sanitize@url \@href}%
\providecommand \@href[1]{\@@startlink{#1}\@@href}%
\providecommand \@@href[1]{\endgroup#1\@@endlink}%
\providecommand \@sanitize@url [0]{\catcode `\\12\catcode `\$12\catcode
  `\&12\catcode `\#12\catcode `\^12\catcode `\_12\catcode `\%12\relax}%
\providecommand \@@startlink[1]{}%
\providecommand \@@endlink[0]{}%
\providecommand \url  [0]{\begingroup\@sanitize@url \@url }%
\providecommand \@url [1]{\endgroup\@href {#1}{\urlprefix }}%
\providecommand \urlprefix  [0]{URL }%
\providecommand \Eprint [0]{\href }%
\providecommand \doibase [0]{http://dx.doi.org/}%
\providecommand \selectlanguage [0]{\@gobble}%
\providecommand \bibinfo  [0]{\@secondoftwo}%
\providecommand \bibfield  [0]{\@secondoftwo}%
\providecommand \translation [1]{[#1]}%
\providecommand \BibitemOpen [0]{}%
\providecommand \bibitemStop [0]{}%
\providecommand \bibitemNoStop [0]{.\EOS\space}%
\providecommand \EOS [0]{\spacefactor3000\relax}%
\providecommand \BibitemShut  [1]{\csname bibitem#1\endcsname}%
\let\auto@bib@innerbib\@empty
\bibitem [{\citenamefont {Gonoskov}\ \emph {et~al.}(2022)\citenamefont
  {Gonoskov}, \citenamefont {Blackburn}, \citenamefont {Marklund},\ and\
  \citenamefont {Bulanov}}]{gonoskov_rmp22}%
  \BibitemOpen
  \bibfield  {author} {\bibinfo {author} {\bibfnamefont {A.}~\bibnamefont
  {Gonoskov}}, \bibinfo {author} {\bibfnamefont {T.}~\bibnamefont {Blackburn}},
  \bibinfo {author} {\bibfnamefont {M.}~\bibnamefont {Marklund}}, \ and\
  \bibinfo {author} {\bibfnamefont {S.}~\bibnamefont {Bulanov}},\ }\href@noop
  {} {\bibfield  {journal} {\bibinfo  {journal} {Reviews of Modern Physics}\
  }\textbf {\bibinfo {volume} {94}},\ \bibinfo {pages} {045001} (\bibinfo
  {year} {2022})}\BibitemShut {NoStop}%
\bibitem [{\citenamefont {Fedotov}\ \emph {et~al.}(2023)\citenamefont
  {Fedotov}, \citenamefont {Ilderton}, \citenamefont {Karbstein}, \citenamefont
  {King}, \citenamefont {Seipt}, \citenamefont {Taya},\ and\ \citenamefont
  {Torgrimsson}}]{fedotov_pr23}%
  \BibitemOpen
  \bibfield  {author} {\bibinfo {author} {\bibfnamefont {A.}~\bibnamefont
  {Fedotov}}, \bibinfo {author} {\bibfnamefont {A.}~\bibnamefont {Ilderton}},
  \bibinfo {author} {\bibfnamefont {F.}~\bibnamefont {Karbstein}}, \bibinfo
  {author} {\bibfnamefont {B.}~\bibnamefont {King}}, \bibinfo {author}
  {\bibfnamefont {D.}~\bibnamefont {Seipt}}, \bibinfo {author} {\bibfnamefont
  {H.}~\bibnamefont {Taya}}, \ and\ \bibinfo {author} {\bibfnamefont
  {G.}~\bibnamefont {Torgrimsson}},\ }\href@noop {} {\bibfield  {journal}
  {\bibinfo  {journal} {Physics Reports}\ }\textbf {\bibinfo {volume} {1010}},\
  \bibinfo {pages} {1} (\bibinfo {year} {2023})}\BibitemShut {NoStop}%
\bibitem [{\citenamefont {Popruzhenko}\ and\ \citenamefont
  {Fedotov}(2023)}]{popruzhenko_usp23}%
  \BibitemOpen
  \bibfield  {author} {\bibinfo {author} {\bibfnamefont {S.~V.}\ \bibnamefont
  {Popruzhenko}}\ and\ \bibinfo {author} {\bibfnamefont {A.~M.}\ \bibnamefont
  {Fedotov}},\ }\href {\doibase 10.3367/UFNe.2023.03.039335} {\bibfield
  {journal} {\bibinfo  {journal} {Phys. Usp.}\ }\textbf {\bibinfo {volume}
  {66}},\ \bibinfo {pages} {460} (\bibinfo {year} {2023})}\BibitemShut
  {NoStop}%
\bibitem [{\citenamefont {Landau}\ and\ \citenamefont
  {Lifshitz}(1967)}]{Landau2}%
  \BibitemOpen
  \bibfield  {author} {\bibinfo {author} {\bibfnamefont {L.}~\bibnamefont
  {Landau}}\ and\ \bibinfo {author} {\bibfnamefont {E.}~\bibnamefont
  {Lifshitz}},\ }\href@noop {} {\emph {\bibinfo {title} {Theoretical physics.
  Field theory}}}\ (\bibinfo  {publisher} {Nauka},\ \bibinfo {year}
  {1967})\BibitemShut {NoStop}%
\bibitem [{\citenamefont {Jackson}(2021)}]{Jackson}%
  \BibitemOpen
  \bibfield  {author} {\bibinfo {author} {\bibfnamefont {J.~D.}\ \bibnamefont
  {Jackson}},\ }\href@noop {} {\emph {\bibinfo {title} {Classical
  electrodynamics}}}\ (\bibinfo  {publisher} {John Wiley \& Sons},\ \bibinfo
  {year} {2021})\BibitemShut {NoStop}%
\bibitem [{\citenamefont {Wistisen}\ \emph {et~al.}(2018)\citenamefont
  {Wistisen}, \citenamefont {Di~Piazza}, \citenamefont {Knudsen},\ and\
  \citenamefont {Uggerh\o{}j}}]{dipiazza_ncomm18}%
  \BibitemOpen
  \bibfield  {author} {\bibinfo {author} {\bibfnamefont {T.~N.}\ \bibnamefont
  {Wistisen}}, \bibinfo {author} {\bibfnamefont {A.}~\bibnamefont {Di~Piazza}},
  \bibinfo {author} {\bibfnamefont {H.~V.}\ \bibnamefont {Knudsen}}, \ and\
  \bibinfo {author} {\bibfnamefont {U.~I.}\ \bibnamefont {Uggerh\o{}j}},\
  }\href {\doibase 10.1038/s41467-018-03165-4} {\bibfield  {journal} {\bibinfo
  {journal} {Nature Communications}\ }\textbf {\bibinfo {volume} {9}},\
  \bibinfo {pages} {795} (\bibinfo {year} {2018})}\BibitemShut {NoStop}%
\bibitem [{\citenamefont {Wistisen}\ \emph {et~al.}(2019)\citenamefont
  {Wistisen}, \citenamefont {Di~Piazza}, \citenamefont {Nielsen}, \citenamefont
  {S\o{}rensen},\ and\ \citenamefont {Uggerh\o{}j}}]{dipiazza_prr19}%
  \BibitemOpen
  \bibfield  {author} {\bibinfo {author} {\bibfnamefont {T.~N.}\ \bibnamefont
  {Wistisen}}, \bibinfo {author} {\bibfnamefont {A.}~\bibnamefont {Di~Piazza}},
  \bibinfo {author} {\bibfnamefont {C.~F.}\ \bibnamefont {Nielsen}}, \bibinfo
  {author} {\bibfnamefont {A.~H.}\ \bibnamefont {S\o{}rensen}}, \ and\ \bibinfo
  {author} {\bibfnamefont {U.~I.}\ \bibnamefont {Uggerh\o{}j}} (\bibinfo
  {collaboration} {CERN NA63}),\ }\href {\doibase
  10.1103/PhysRevResearch.1.033014} {\bibfield  {journal} {\bibinfo  {journal}
  {Physical Review Research}\ }\textbf {\bibinfo {volume} {1}},\ \bibinfo
  {pages} {033014} (\bibinfo {year} {2019})}\BibitemShut {NoStop}%
\bibitem [{\citenamefont {{Di Piazza}}\ \emph {et~al.}(2017)\citenamefont {{Di
  Piazza}}, \citenamefont {Wistisen},\ and\ \citenamefont
  {Uggerhøj}}]{dipiazza_plb17}%
  \BibitemOpen
  \bibfield  {author} {\bibinfo {author} {\bibfnamefont {A.}~\bibnamefont {{Di
  Piazza}}}, \bibinfo {author} {\bibfnamefont {T.~N.}\ \bibnamefont
  {Wistisen}}, \ and\ \bibinfo {author} {\bibfnamefont {U.~I.}\ \bibnamefont
  {Uggerhøj}},\ }\href {\doibase
  https://doi.org/10.1016/j.physletb.2016.10.083} {\bibfield  {journal}
  {\bibinfo  {journal} {Physics Letters B}\ }\textbf {\bibinfo {volume}
  {765}},\ \bibinfo {pages} {1} (\bibinfo {year} {2017})}\BibitemShut {NoStop}%
\bibitem [{\citenamefont {Khokonov}(2019)}]{khokonov_plb19}%
  \BibitemOpen
  \bibfield  {author} {\bibinfo {author} {\bibfnamefont {M.}~\bibnamefont
  {Khokonov}},\ }\href {\doibase
  https://doi.org/10.1016/j.physletb.2019.02.034} {\bibfield  {journal}
  {\bibinfo  {journal} {Physics Letters B}\ }\textbf {\bibinfo {volume}
  {791}},\ \bibinfo {pages} {281} (\bibinfo {year} {2019})}\BibitemShut
  {NoStop}%
\bibitem [{\citenamefont {Cole}\ \emph {et~al.}(2018)\citenamefont {Cole},
  \citenamefont {Behm}, \citenamefont {Gerstmayr}, \citenamefont {Blackburn},
  \citenamefont {Wood}, \citenamefont {Baird}, \citenamefont {Duff},
  \citenamefont {Harvey}, \citenamefont {Ilderton}, \citenamefont {Joglekar}
  \emph {et~al.}}]{cole_prx18}%
  \BibitemOpen
  \bibfield  {author} {\bibinfo {author} {\bibfnamefont {J.}~\bibnamefont
  {Cole}}, \bibinfo {author} {\bibfnamefont {K.}~\bibnamefont {Behm}}, \bibinfo
  {author} {\bibfnamefont {E.}~\bibnamefont {Gerstmayr}}, \bibinfo {author}
  {\bibfnamefont {T.}~\bibnamefont {Blackburn}}, \bibinfo {author}
  {\bibfnamefont {J.}~\bibnamefont {Wood}}, \bibinfo {author} {\bibfnamefont
  {C.}~\bibnamefont {Baird}}, \bibinfo {author} {\bibfnamefont {M.~J.}\
  \bibnamefont {Duff}}, \bibinfo {author} {\bibfnamefont {C.}~\bibnamefont
  {Harvey}}, \bibinfo {author} {\bibfnamefont {A.}~\bibnamefont {Ilderton}},
  \bibinfo {author} {\bibfnamefont {A.}~\bibnamefont {Joglekar}},  \emph
  {et~al.},\ }\href@noop {} {\bibfield  {journal} {\bibinfo  {journal}
  {Physical Review X}\ }\textbf {\bibinfo {volume} {8}},\ \bibinfo {pages}
  {011020} (\bibinfo {year} {2018})}\BibitemShut {NoStop}%
\bibitem [{\citenamefont {Poder}\ \emph {et~al.}(2018)\citenamefont {Poder},
  \citenamefont {Tamburini}, \citenamefont {Sarri}, \citenamefont {Di~Piazza},
  \citenamefont {Kuschel}, \citenamefont {Baird}, \citenamefont {Behm},
  \citenamefont {Bohlen}, \citenamefont {Cole}, \citenamefont {Corvan} \emph
  {et~al.}}]{poder_prx18}%
  \BibitemOpen
  \bibfield  {author} {\bibinfo {author} {\bibfnamefont {K.}~\bibnamefont
  {Poder}}, \bibinfo {author} {\bibfnamefont {M.}~\bibnamefont {Tamburini}},
  \bibinfo {author} {\bibfnamefont {G.}~\bibnamefont {Sarri}}, \bibinfo
  {author} {\bibfnamefont {A.}~\bibnamefont {Di~Piazza}}, \bibinfo {author}
  {\bibfnamefont {S.}~\bibnamefont {Kuschel}}, \bibinfo {author} {\bibfnamefont
  {C.}~\bibnamefont {Baird}}, \bibinfo {author} {\bibfnamefont
  {K.}~\bibnamefont {Behm}}, \bibinfo {author} {\bibfnamefont {S.}~\bibnamefont
  {Bohlen}}, \bibinfo {author} {\bibfnamefont {J.}~\bibnamefont {Cole}},
  \bibinfo {author} {\bibfnamefont {D.}~\bibnamefont {Corvan}},  \emph
  {et~al.},\ }\href@noop {} {\bibfield  {journal} {\bibinfo  {journal}
  {Physical Review X}\ }\textbf {\bibinfo {volume} {8}},\ \bibinfo {pages}
  {031004} (\bibinfo {year} {2018})}\BibitemShut {NoStop}%
\bibitem [{\citenamefont {Macchi}(2018)}]{macchi_phys18}%
  \BibitemOpen
  \bibfield  {author} {\bibinfo {author} {\bibfnamefont {A.}~\bibnamefont
  {Macchi}},\ }\href@noop {} {\bibfield  {journal} {\bibinfo  {journal}
  {Physics}\ }\textbf {\bibinfo {volume} {11}},\ \bibinfo {pages} {13}
  (\bibinfo {year} {2018})}\BibitemShut {NoStop}%
\bibitem [{\citenamefont {Gelfer}\ \emph {et~al.}(2015)\citenamefont {Gelfer},
  \citenamefont {Mironov}, \citenamefont {Fedotov}, \citenamefont {Bashmakov},
  \citenamefont {Nerush}, \citenamefont {Kostyukov},\ and\ \citenamefont
  {Narozhny}}]{gelfer_pra15}%
  \BibitemOpen
  \bibfield  {author} {\bibinfo {author} {\bibfnamefont {E.}~\bibnamefont
  {Gelfer}}, \bibinfo {author} {\bibfnamefont {A.}~\bibnamefont {Mironov}},
  \bibinfo {author} {\bibfnamefont {A.}~\bibnamefont {Fedotov}}, \bibinfo
  {author} {\bibfnamefont {V.}~\bibnamefont {Bashmakov}}, \bibinfo {author}
  {\bibfnamefont {E.}~\bibnamefont {Nerush}}, \bibinfo {author} {\bibfnamefont
  {I.~Y.}\ \bibnamefont {Kostyukov}}, \ and\ \bibinfo {author} {\bibfnamefont
  {N.}~\bibnamefont {Narozhny}},\ }\href@noop {} {\bibfield  {journal}
  {\bibinfo  {journal} {Physical Review A}\ }\textbf {\bibinfo {volume} {92}},\
  \bibinfo {pages} {022113} (\bibinfo {year} {2015})}\BibitemShut {NoStop}%
\bibitem [{\citenamefont {Nerush}\ \emph {et~al.}(2011)\citenamefont {Nerush},
  \citenamefont {Kostyukov}, \citenamefont {Fedotov}, \citenamefont {Narozhny},
  \citenamefont {Elkina},\ and\ \citenamefont {Ruhl}}]{kostyukov_prl11}%
  \BibitemOpen
  \bibfield  {author} {\bibinfo {author} {\bibfnamefont {E.~N.}\ \bibnamefont
  {Nerush}}, \bibinfo {author} {\bibfnamefont {I.~Y.}\ \bibnamefont
  {Kostyukov}}, \bibinfo {author} {\bibfnamefont {A.~M.}\ \bibnamefont
  {Fedotov}}, \bibinfo {author} {\bibfnamefont {N.~B.}\ \bibnamefont
  {Narozhny}}, \bibinfo {author} {\bibfnamefont {N.~V.}\ \bibnamefont
  {Elkina}}, \ and\ \bibinfo {author} {\bibfnamefont {H.}~\bibnamefont
  {Ruhl}},\ }\href {\doibase 10.1103/PhysRevLett.106.035001} {\bibfield
  {journal} {\bibinfo  {journal} {Physical Review Letters}\ }\textbf {\bibinfo
  {volume} {106}},\ \bibinfo {pages} {035001} (\bibinfo {year}
  {2011})}\BibitemShut {NoStop}%
\bibitem [{\citenamefont {Mourou}\ \emph {et~al.}(2006)\citenamefont {Mourou},
  \citenamefont {Tajima},\ and\ \citenamefont {Bulanov}}]{bulanov_rmp06}%
  \BibitemOpen
  \bibfield  {author} {\bibinfo {author} {\bibfnamefont {G.~A.}\ \bibnamefont
  {Mourou}}, \bibinfo {author} {\bibfnamefont {T.}~\bibnamefont {Tajima}}, \
  and\ \bibinfo {author} {\bibfnamefont {S.~V.}\ \bibnamefont {Bulanov}},\
  }\href@noop {} {\bibfield  {journal} {\bibinfo  {journal} {Reviews of modern
  physics}\ }\textbf {\bibinfo {volume} {78}},\ \bibinfo {pages} {309}
  (\bibinfo {year} {2006})}\BibitemShut {NoStop}%
\bibitem [{\citenamefont {Blackburn}(2020)}]{blackburn_rmpp20}%
  \BibitemOpen
  \bibfield  {author} {\bibinfo {author} {\bibfnamefont {T.}~\bibnamefont
  {Blackburn}},\ }\href@noop {} {\bibfield  {journal} {\bibinfo  {journal}
  {Reviews of Modern Plasma Physics}\ }\textbf {\bibinfo {volume} {4}},\
  \bibinfo {pages} {5} (\bibinfo {year} {2020})}\BibitemShut {NoStop}%
\bibitem [{\citenamefont {Pitaevskii}(1961)}]{pitaevskii_jetp61}%
  \BibitemOpen
  \bibfield  {author} {\bibinfo {author} {\bibfnamefont {L.}~\bibnamefont
  {Pitaevskii}},\ }\href@noop {} {\bibfield  {journal} {\bibinfo  {journal}
  {Sov. Phys. JETP}\ }\textbf {\bibinfo {volume} {12}},\ \bibinfo {pages}
  {1008} (\bibinfo {year} {1961})}\BibitemShut {NoStop}%
\bibitem [{\citenamefont {Pershan}(1963)}]{pershan_pr63}%
  \BibitemOpen
  \bibfield  {author} {\bibinfo {author} {\bibfnamefont {P.}~\bibnamefont
  {Pershan}},\ }\href@noop {} {\bibfield  {journal} {\bibinfo  {journal}
  {Physical Review}\ }\textbf {\bibinfo {volume} {130}},\ \bibinfo {pages}
  {919} (\bibinfo {year} {1963})}\BibitemShut {NoStop}%
\bibitem [{\citenamefont {Van~der Ziel}\ \emph {et~al.}(1965)\citenamefont
  {Van~der Ziel}, \citenamefont {Pershan},\ and\ \citenamefont
  {Malmstrom}}]{van_prl65}%
  \BibitemOpen
  \bibfield  {author} {\bibinfo {author} {\bibfnamefont {J.}~\bibnamefont
  {Van~der Ziel}}, \bibinfo {author} {\bibfnamefont {P.~S.}\ \bibnamefont
  {Pershan}}, \ and\ \bibinfo {author} {\bibfnamefont {L.}~\bibnamefont
  {Malmstrom}},\ }\href@noop {} {\bibfield  {journal} {\bibinfo  {journal}
  {Physical Review Letters}\ }\textbf {\bibinfo {volume} {15}},\ \bibinfo
  {pages} {190} (\bibinfo {year} {1965})}\BibitemShut {NoStop}%
\bibitem [{\citenamefont {Liseykina}\ \emph {et~al.}(2016)\citenamefont
  {Liseykina}, \citenamefont {Popruzhenko},\ and\ \citenamefont
  {Macchi}}]{macchi_njp16}%
  \BibitemOpen
  \bibfield  {author} {\bibinfo {author} {\bibfnamefont {T.~V.}\ \bibnamefont
  {Liseykina}}, \bibinfo {author} {\bibfnamefont {S.~V.}\ \bibnamefont
  {Popruzhenko}}, \ and\ \bibinfo {author} {\bibfnamefont {A.}~\bibnamefont
  {Macchi}},\ }\href {\doibase 10.1088/1367-2630/18/7/072001} {\bibfield
  {journal} {\bibinfo  {journal} {New Journal of Physics}\ }\textbf {\bibinfo
  {volume} {18}},\ \bibinfo {pages} {072001} (\bibinfo {year}
  {2016})}\BibitemShut {NoStop}%
\bibitem [{\citenamefont {Popruzhenko}\ \emph {et~al.}(2019)\citenamefont
  {Popruzhenko}, \citenamefont {Liseykina},\ and\ \citenamefont
  {Macchi}}]{poprz_njp19}%
  \BibitemOpen
  \bibfield  {author} {\bibinfo {author} {\bibfnamefont {S.~V.}\ \bibnamefont
  {Popruzhenko}}, \bibinfo {author} {\bibfnamefont {T.~V.}\ \bibnamefont
  {Liseykina}}, \ and\ \bibinfo {author} {\bibfnamefont {A.}~\bibnamefont
  {Macchi}},\ }\href {\doibase 10.1088/1367-2630/ab0119} {\bibfield  {journal}
  {\bibinfo  {journal} {New Journal of Physics}\ }\textbf {\bibinfo {volume}
  {21}},\ \bibinfo {pages} {033009} (\bibinfo {year} {2019})}\BibitemShut
  {NoStop}%
\bibitem [{\citenamefont {Liseykina}\ \emph {et~al.}(2021)\citenamefont
  {Liseykina}, \citenamefont {Macchi},\ and\ \citenamefont
  {Popruzhenko}}]{poprz_ejpp21}%
  \BibitemOpen
  \bibfield  {author} {\bibinfo {author} {\bibfnamefont {T.~V.}\ \bibnamefont
  {Liseykina}}, \bibinfo {author} {\bibfnamefont {A.}~\bibnamefont {Macchi}}, \
  and\ \bibinfo {author} {\bibfnamefont {S.~V.}\ \bibnamefont {Popruzhenko}},\
  }\href {\doibase 10.1140/epjp/s13360-020-01030-2} {\bibfield  {journal}
  {\bibinfo  {journal} {The European Physical Journal Plus}\ }\textbf {\bibinfo
  {volume} {136}},\ \bibinfo {pages} {170} (\bibinfo {year}
  {2021})}\BibitemShut {NoStop}%
\bibitem [{\citenamefont {Papadopoulos}\ \emph {et~al.}(2024)\citenamefont
  {Papadopoulos}, \citenamefont {Ayoul}, \citenamefont {Beluze}, \citenamefont
  {Gobert}, \citenamefont {Mataja}, \citenamefont {Fr{\'e}neaux}, \citenamefont
  {Lebas}, \citenamefont {Chabanis}, \citenamefont {Dumergue}, \citenamefont
  {Audebert},\ and\ \citenamefont {Mathieu}}]{apollon}%
  \BibitemOpen
  \bibfield  {author} {\bibinfo {author} {\bibfnamefont {D.}~\bibnamefont
  {Papadopoulos}}, \bibinfo {author} {\bibfnamefont {Y.}~\bibnamefont {Ayoul}},
  \bibinfo {author} {\bibfnamefont {A.}~\bibnamefont {Beluze}}, \bibinfo
  {author} {\bibfnamefont {F.}~\bibnamefont {Gobert}}, \bibinfo {author}
  {\bibfnamefont {D.}~\bibnamefont {Mataja}}, \bibinfo {author} {\bibfnamefont
  {A.}~\bibnamefont {Fr{\'e}neaux}}, \bibinfo {author} {\bibfnamefont
  {N.}~\bibnamefont {Lebas}}, \bibinfo {author} {\bibfnamefont
  {M.}~\bibnamefont {Chabanis}}, \bibinfo {author} {\bibfnamefont
  {M.}~\bibnamefont {Dumergue}}, \bibinfo {author} {\bibfnamefont
  {P.}~\bibnamefont {Audebert}}, \ and\ \bibinfo {author} {\bibfnamefont
  {F.}~\bibnamefont {Mathieu}},\ }in\ \href@noop {} {\emph {\bibinfo
  {booktitle} {High Intensity Lasers and High Field Phenomena}}}\ (\bibinfo
  {organization} {Optica Publishing Group},\ \bibinfo {year} {2024})\ pp.\
  \bibinfo {pages} {HTu2B--2}\BibitemShut {NoStop}%
\bibitem [{\citenamefont {Yoon}\ \emph {et~al.}(2021)\citenamefont {Yoon},
  \citenamefont {Kim}, \citenamefont {Choi}, \citenamefont {Sung},
  \citenamefont {Lee}, \citenamefont {Lee},\ and\ \citenamefont
  {Nam}}]{corels}%
  \BibitemOpen
  \bibfield  {author} {\bibinfo {author} {\bibfnamefont {J.~W.}\ \bibnamefont
  {Yoon}}, \bibinfo {author} {\bibfnamefont {Y.~G.}\ \bibnamefont {Kim}},
  \bibinfo {author} {\bibfnamefont {I.~W.}\ \bibnamefont {Choi}}, \bibinfo
  {author} {\bibfnamefont {J.~H.}\ \bibnamefont {Sung}}, \bibinfo {author}
  {\bibfnamefont {H.~W.}\ \bibnamefont {Lee}}, \bibinfo {author} {\bibfnamefont
  {S.~K.}\ \bibnamefont {Lee}}, \ and\ \bibinfo {author} {\bibfnamefont
  {C.~H.}\ \bibnamefont {Nam}},\ }\href {\doibase 10.1364/OPTICA.420520}
  {\bibfield  {journal} {\bibinfo  {journal} {Optica}\ }\textbf {\bibinfo
  {volume} {8}},\ \bibinfo {pages} {630} (\bibinfo {year} {2021})}\BibitemShut
  {NoStop}%
\bibitem [{https://www.eli-beams.eu()}]{ELI-BL}%
  \BibitemOpen
  https://www.eli-beams.eu,\ \href@noop {} {}\BibitemShut {NoStop}%
\bibitem [{\citenamefont {Doria}\ \emph {et~al.}(2020)\citenamefont {Doria},
  \citenamefont {Cernaianu}, \citenamefont {Ghenuche}, \citenamefont {Stutman},
  \citenamefont {Tanaka}, \citenamefont {Ticos},\ and\ \citenamefont
  {Ur}}]{ELI-NP}%
  \BibitemOpen
  \bibfield  {author} {\bibinfo {author} {\bibfnamefont {D.}~\bibnamefont
  {Doria}}, \bibinfo {author} {\bibfnamefont {M.}~\bibnamefont {Cernaianu}},
  \bibinfo {author} {\bibfnamefont {P.}~\bibnamefont {Ghenuche}}, \bibinfo
  {author} {\bibfnamefont {D.}~\bibnamefont {Stutman}}, \bibinfo {author}
  {\bibfnamefont {K.}~\bibnamefont {Tanaka}}, \bibinfo {author} {\bibfnamefont
  {C.}~\bibnamefont {Ticos}}, \ and\ \bibinfo {author} {\bibfnamefont
  {C.}~\bibnamefont {Ur}},\ }\href@noop {} {\bibfield  {journal} {\bibinfo
  {journal} {Journal of Instrumentation}\ }\textbf {\bibinfo {volume} {15}},\
  \bibinfo {pages} {C09053} (\bibinfo {year} {2020})}\BibitemShut {NoStop}%
\bibitem [{\citenamefont {Zhang}\ \emph {et~al.}(2020)\citenamefont {Zhang},
  \citenamefont {Wu}, \citenamefont {Hu}, \citenamefont {Yang}, \citenamefont
  {Gui}, \citenamefont {Ji}, \citenamefont {Liu}, \citenamefont {Wang},
  \citenamefont {Liu}, \citenamefont {Lu}, \citenamefont {Xu}, \citenamefont
  {Leng}, \citenamefont {Li},\ and\ \citenamefont {Xu}}]{SULF}%
  \BibitemOpen
  \bibfield  {author} {\bibinfo {author} {\bibfnamefont {Z.}~\bibnamefont
  {Zhang}}, \bibinfo {author} {\bibfnamefont {F.}~\bibnamefont {Wu}}, \bibinfo
  {author} {\bibfnamefont {J.}~\bibnamefont {Hu}}, \bibinfo {author}
  {\bibfnamefont {X.}~\bibnamefont {Yang}}, \bibinfo {author} {\bibfnamefont
  {J.}~\bibnamefont {Gui}}, \bibinfo {author} {\bibfnamefont {P.}~\bibnamefont
  {Ji}}, \bibinfo {author} {\bibfnamefont {X.}~\bibnamefont {Liu}}, \bibinfo
  {author} {\bibfnamefont {C.}~\bibnamefont {Wang}}, \bibinfo {author}
  {\bibfnamefont {Y.}~\bibnamefont {Liu}}, \bibinfo {author} {\bibfnamefont
  {X.}~\bibnamefont {Lu}}, \bibinfo {author} {\bibfnamefont {Y.}~\bibnamefont
  {Xu}}, \bibinfo {author} {\bibfnamefont {Y.}~\bibnamefont {Leng}}, \bibinfo
  {author} {\bibfnamefont {R.}~\bibnamefont {Li}}, \ and\ \bibinfo {author}
  {\bibfnamefont {Z.}~\bibnamefont {Xu}},\ }\href@noop {} {\bibfield  {journal}
  {\bibinfo  {journal} {High Power Laser Science and Engineering}\ }\textbf
  {\bibinfo {volume} {8}},\ \bibinfo {pages} {e4} (\bibinfo {year}
  {2020})}\BibitemShut {NoStop}%
\bibitem [{\citenamefont {Kostyukov}\ \emph {et~al.}(2023)\citenamefont
  {Kostyukov}, \citenamefont {Khazanov}, \citenamefont {Shaikin}, \citenamefont
  {Litvak},\ and\ \citenamefont {Sergeev}}]{XCELS}%
  \BibitemOpen
  \bibfield  {author} {\bibinfo {author} {\bibfnamefont {I.~Y.}\ \bibnamefont
  {Kostyukov}}, \bibinfo {author} {\bibfnamefont {E.}~\bibnamefont {Khazanov}},
  \bibinfo {author} {\bibfnamefont {A.}~\bibnamefont {Shaikin}}, \bibinfo
  {author} {\bibfnamefont {A.}~\bibnamefont {Litvak}}, \ and\ \bibinfo {author}
  {\bibfnamefont {A.}~\bibnamefont {Sergeev}},\ }\href@noop {} {\bibfield
  {journal} {\bibinfo  {journal} {Bulletin of the Lebedev Physics Institute}\
  }\textbf {\bibinfo {volume} {50}},\ \bibinfo {pages} {S635} (\bibinfo {year}
  {2023})}\BibitemShut {NoStop}%
\bibitem [{\citenamefont {Vshivkov}\ and\ \citenamefont
  {Dudnikova}(2001)}]{umka}%
  \BibitemOpen
  \bibfield  {author} {\bibinfo {author} {\bibfnamefont {V.~A.}\ \bibnamefont
  {Vshivkov}}\ and\ \bibinfo {author} {\bibfnamefont {G.~I.}\ \bibnamefont
  {Dudnikova}},\ }\href {https://dx.doi.org/10.1088/1367-2630/ab0119}
  {\bibfield  {journal} {\bibinfo  {journal} {Computational technologies}\
  }\textbf {\bibinfo {volume} {6}},\ \bibinfo {pages} {47} (\bibinfo {year}
  {2001})}\BibitemShut {NoStop}%
\bibitem [{\citenamefont {Vshivkov}\ \emph {et~al.}(1998)\citenamefont
  {Vshivkov}, \citenamefont {Naumova}, \citenamefont {Pegoraro},\ and\
  \citenamefont {Bulanov}}]{umka_1998}%
  \BibitemOpen
  \bibfield  {author} {\bibinfo {author} {\bibfnamefont {V.~A.}\ \bibnamefont
  {Vshivkov}}, \bibinfo {author} {\bibfnamefont {N.~M.}\ \bibnamefont
  {Naumova}}, \bibinfo {author} {\bibfnamefont {F.}~\bibnamefont {Pegoraro}}, \
  and\ \bibinfo {author} {\bibfnamefont {S.~V.}\ \bibnamefont {Bulanov}},\
  }\href {\doibase 10.1063/1.872961} {\bibfield  {journal} {\bibinfo  {journal}
  {Physics of Plasmas}\ }\textbf {\bibinfo {volume} {5}},\ \bibinfo {pages}
  {2727} (\bibinfo {year} {1998})}\BibitemShut {NoStop}%
\bibitem [{\citenamefont {Liseykina}\ \emph {et~al.}(2020)\citenamefont
  {Liseykina}, \citenamefont {Bauer}, \citenamefont {Popruzhenko},\ and\
  \citenamefont {Macchi}}]{NIC_2020}%
  \BibitemOpen
  \bibfield  {author} {\bibinfo {author} {\bibfnamefont {T.}~\bibnamefont
  {Liseykina}}, \bibinfo {author} {\bibfnamefont {D.}~\bibnamefont {Bauer}},
  \bibinfo {author} {\bibfnamefont {S.}~\bibnamefont {Popruzhenko}}, \ and\
  \bibinfo {author} {\bibfnamefont {A.}~\bibnamefont {Macchi}},\ }in\ \href
  {http://hdl.handle.net/2128/24562} {\emph {\bibinfo {booktitle} {Proceedings
  of the International Conference on Example}}},\ \bibinfo {series} {NIC
  Series}, Vol.~\bibinfo {volume} {50},\ \bibinfo {editor} {edited by\ \bibinfo
  {editor} {\bibfnamefont {M.}~\bibnamefont {Müller}}, \bibinfo {editor}
  {\bibfnamefont {K.}~\bibnamefont {Binder}}, \ and\ \bibinfo {editor}
  {\bibfnamefont {A.}~\bibnamefont {Trautmann}}},\ \bibinfo {organization} {NIC
  Symposium, Jülich (Germany), 27 Feb 2020 - 28 Feb 2020}\ (\bibinfo
  {publisher} {Forschungszentrum Jülich GmbH Zentralbibliothek, Verlag},\
  \bibinfo {address} {Jülich},\ \bibinfo {year} {2020})\ pp.\ \bibinfo {pages}
  {405--414}\BibitemShut {NoStop}%
\bibitem [{\citenamefont {Derouillat}\ \emph {et~al.}(2018)\citenamefont
  {Derouillat}, \citenamefont {Beck}, \citenamefont {Pérez}, \citenamefont
  {Vinci}, \citenamefont {Chiaramello}, \citenamefont {Grassi}, \citenamefont
  {Flé}, \citenamefont {Bouchard}, \citenamefont {Plotnikov}, \citenamefont
  {Aunai}, \citenamefont {Dargent}, \citenamefont {Riconda},\ and\
  \citenamefont {Grech}}]{smilei}%
  \BibitemOpen
  \bibfield  {author} {\bibinfo {author} {\bibfnamefont {J.}~\bibnamefont
  {Derouillat}}, \bibinfo {author} {\bibfnamefont {A.}~\bibnamefont {Beck}},
  \bibinfo {author} {\bibfnamefont {F.}~\bibnamefont {Pérez}}, \bibinfo
  {author} {\bibfnamefont {T.}~\bibnamefont {Vinci}}, \bibinfo {author}
  {\bibfnamefont {M.}~\bibnamefont {Chiaramello}}, \bibinfo {author}
  {\bibfnamefont {A.}~\bibnamefont {Grassi}}, \bibinfo {author} {\bibfnamefont
  {M.}~\bibnamefont {Flé}}, \bibinfo {author} {\bibfnamefont {G.}~\bibnamefont
  {Bouchard}}, \bibinfo {author} {\bibfnamefont {I.}~\bibnamefont {Plotnikov}},
  \bibinfo {author} {\bibfnamefont {N.}~\bibnamefont {Aunai}}, \bibinfo
  {author} {\bibfnamefont {J.}~\bibnamefont {Dargent}}, \bibinfo {author}
  {\bibfnamefont {C.}~\bibnamefont {Riconda}}, \ and\ \bibinfo {author}
  {\bibfnamefont {M.}~\bibnamefont {Grech}},\ }\href {\doibase
  https://doi.org/10.1016/j.cpc.2017.09.024} {\bibfield  {journal} {\bibinfo
  {journal} {Computer Physics Communications}\ }\textbf {\bibinfo {volume}
  {222}},\ \bibinfo {pages} {351} (\bibinfo {year} {2018})}\BibitemShut
  {NoStop}%
\bibitem [{\citenamefont {Ritus}(1985)}]{ritus_jslr85}%
  \BibitemOpen
  \bibfield  {author} {\bibinfo {author} {\bibfnamefont {V.~I.}\ \bibnamefont
  {Ritus}},\ }\href@noop {} {\bibfield  {journal} {\bibinfo  {journal} {Journal
  of Soviet Laser Research}\ }\textbf {\bibinfo {volume} {6}},\ \bibinfo
  {pages} {497} (\bibinfo {year} {1985})}\BibitemShut {NoStop}%
\bibitem [{\citenamefont {Kirk}\ \emph {et~al.}(2009)\citenamefont {Kirk},
  \citenamefont {Bell},\ and\ \citenamefont {Arka}}]{kirk_ppcf09}%
  \BibitemOpen
  \bibfield  {author} {\bibinfo {author} {\bibfnamefont {J.~G.}\ \bibnamefont
  {Kirk}}, \bibinfo {author} {\bibfnamefont {A.}~\bibnamefont {Bell}}, \ and\
  \bibinfo {author} {\bibfnamefont {I.}~\bibnamefont {Arka}},\ }\href@noop {}
  {\bibfield  {journal} {\bibinfo  {journal} {Plasma Physics and Controlled
  Fusion}\ }\textbf {\bibinfo {volume} {51}},\ \bibinfo {pages} {085008}
  (\bibinfo {year} {2009})}\BibitemShut {NoStop}%
\bibitem [{\citenamefont {Liseykina}\ \emph {et~al.}(2023)\citenamefont
  {Liseykina}, \citenamefont {Peganov},\ and\ \citenamefont
  {Popruzhenko}}]{liseykina_bull23}%
  \BibitemOpen
  \bibfield  {author} {\bibinfo {author} {\bibfnamefont {T.~V.}\ \bibnamefont
  {Liseykina}}, \bibinfo {author} {\bibfnamefont {E.~E.}\ \bibnamefont
  {Peganov}}, \ and\ \bibinfo {author} {\bibfnamefont {S.~V.}\ \bibnamefont
  {Popruzhenko}},\ }\href {\doibase 10.3103/S1068335623180082} {\bibfield
  {journal} {\bibinfo  {journal} {Bull. Lebedev Phys. Inst.}\ }\textbf
  {\bibinfo {volume} {50}},\ \bibinfo {pages} {S700} (\bibinfo {year}
  {2023})}\BibitemShut {NoStop}%
\bibitem [{\citenamefont {Yee}(1966)}]{Yee}%
  \BibitemOpen
  \bibfield  {author} {\bibinfo {author} {\bibfnamefont {K.}~\bibnamefont
  {Yee}},\ }\href {\doibase 10.1109/TAP.1966.1138693} {\bibfield  {journal}
  {\bibinfo  {journal} {IEEE Transactions on Antennas and Propagation}\
  }\textbf {\bibinfo {volume} {14}},\ \bibinfo {pages} {302} (\bibinfo {year}
  {1966})}\BibitemShut {NoStop}%
\bibitem [{\citenamefont {LANGDON}\ and\ \citenamefont
  {LASINSKI}(1976)}]{Langdon1976}%
  \BibitemOpen
  \bibfield  {author} {\bibinfo {author} {\bibfnamefont {A.~B.}\ \bibnamefont
  {LANGDON}}\ and\ \bibinfo {author} {\bibfnamefont {B.~F.}\ \bibnamefont
  {LASINSKI}},\ }in\ \href {\doibase
  https://doi.org/10.1016/B978-0-12-460816-0.50014-2} {\emph {\bibinfo
  {booktitle} {Controlled Fusion}}},\ \bibinfo {series} {Methods in
  Computational Physics: Advances in Research and Applications}, Vol.~\bibinfo
  {volume} {16},\ \bibinfo {editor} {edited by\ \bibinfo {editor}
  {\bibfnamefont {J.}~\bibnamefont {Killeen}}}\ (\bibinfo  {publisher}
  {Elsevier},\ \bibinfo {year} {1976})\ pp.\ \bibinfo {pages}
  {327--366}\BibitemShut {NoStop}%
\bibitem [{\citenamefont {Taflove}\ and\ \citenamefont
  {Hagness}(2005)}]{Taflove2005}%
  \BibitemOpen
  \bibfield  {author} {\bibinfo {author} {\bibfnamefont {A.}~\bibnamefont
  {Taflove}}\ and\ \bibinfo {author} {\bibfnamefont {S.~C.}\ \bibnamefont
  {Hagness}},\ }\href@noop {} {\emph {\bibinfo {title} {Computational
  electrodynamics: the finite-difference time-domain method}}},\ \bibinfo
  {edition} {3rd}\ ed.\ (\bibinfo  {publisher} {Artech House},\ \bibinfo
  {address} {Norwood},\ \bibinfo {year} {2005})\BibitemShut {NoStop}%
\bibitem [{\citenamefont {Blackburn}(2024)}]{blackburn_pra24}%
  \BibitemOpen
  \bibfield  {author} {\bibinfo {author} {\bibfnamefont {T.}~\bibnamefont
  {Blackburn}},\ }\href@noop {} {\bibfield  {journal} {\bibinfo  {journal}
  {Physical Review A}\ }\textbf {\bibinfo {volume} {109}},\ \bibinfo {pages}
  {022234} (\bibinfo {year} {2024})}\BibitemShut {NoStop}%
\bibitem [{\citenamefont {Ritus}(1979)}]{ritus}%
  \BibitemOpen
  \bibfield  {author} {\bibinfo {author} {\bibfnamefont {V.~I.}\ \bibnamefont
  {Ritus}},\ }\href@noop {} {\bibfield  {journal} {\bibinfo  {journal} {Moscow
  Izdatel Nauka AN SSR Fizicheskii Institut Trudy}\ }\textbf {\bibinfo {volume}
  {111}},\ \bibinfo {pages} {5} (\bibinfo {year} {1979})}\BibitemShut {NoStop}%
\bibitem [{\citenamefont {Thomas}\ \emph {et~al.}(2012)\citenamefont {Thomas},
  \citenamefont {Ridgers}, \citenamefont {Bulanov}, \citenamefont {Griffin},\
  and\ \citenamefont {Mangles}}]{thomas-prx12}%
  \BibitemOpen
  \bibfield  {author} {\bibinfo {author} {\bibfnamefont {A.~G.~R.}\
  \bibnamefont {Thomas}}, \bibinfo {author} {\bibfnamefont {C.~P.}\
  \bibnamefont {Ridgers}}, \bibinfo {author} {\bibfnamefont {S.~S.}\
  \bibnamefont {Bulanov}}, \bibinfo {author} {\bibfnamefont {B.~J.}\
  \bibnamefont {Griffin}}, \ and\ \bibinfo {author} {\bibfnamefont {S.~P.~D.}\
  \bibnamefont {Mangles}},\ }\href {\doibase 10.1103/PhysRevX.2.041004}
  {\bibfield  {journal} {\bibinfo  {journal} {Physical Review X}\ }\textbf
  {\bibinfo {volume} {2}},\ \bibinfo {pages} {041004} (\bibinfo {year}
  {2012})}\BibitemShut {NoStop}%
\bibitem [{\citenamefont {Tamburini}\ \emph {et~al.}(2010)\citenamefont
  {Tamburini}, \citenamefont {Pegoraro}, \citenamefont {Piazza}, \citenamefont
  {Keitel},\ and\ \citenamefont {Macchi}}]{Tamburini_2010}%
  \BibitemOpen
  \bibfield  {author} {\bibinfo {author} {\bibfnamefont {M.}~\bibnamefont
  {Tamburini}}, \bibinfo {author} {\bibfnamefont {F.}~\bibnamefont {Pegoraro}},
  \bibinfo {author} {\bibfnamefont {A.~D.}\ \bibnamefont {Piazza}}, \bibinfo
  {author} {\bibfnamefont {C.~H.}\ \bibnamefont {Keitel}}, \ and\ \bibinfo
  {author} {\bibfnamefont {A.}~\bibnamefont {Macchi}},\ }\href {\doibase
  10.1088/1367-2630/12/12/123005} {\bibfield  {journal} {\bibinfo  {journal}
  {New Journal of Physics}\ }\textbf {\bibinfo {volume} {12}},\ \bibinfo
  {pages} {123005} (\bibinfo {year} {2010})}\BibitemShut {NoStop}%
\bibitem [{\citenamefont {Gelfer}\ \emph
  {et~al.}(2024{\natexlab{a}})\citenamefont {Gelfer}, \citenamefont {Fedotov},
  \citenamefont {Klimo},\ and\ \citenamefont {Weber}}]{gelfer_mre24}%
  \BibitemOpen
  \bibfield  {author} {\bibinfo {author} {\bibfnamefont {E.}~\bibnamefont
  {Gelfer}}, \bibinfo {author} {\bibfnamefont {A.}~\bibnamefont {Fedotov}},
  \bibinfo {author} {\bibfnamefont {O.}~\bibnamefont {Klimo}}, \ and\ \bibinfo
  {author} {\bibfnamefont {S.}~\bibnamefont {Weber}},\ }\href@noop {}
  {\bibfield  {journal} {\bibinfo  {journal} {Matter and Radiation at
  Extremes}\ }\textbf {\bibinfo {volume} {9}} (\bibinfo {year}
  {2024}{\natexlab{a}})}\BibitemShut {NoStop}%
\bibitem [{\citenamefont {Gelfer}\ \emph
  {et~al.}(2024{\natexlab{b}})\citenamefont {Gelfer}, \citenamefont {Fedotov},
  \citenamefont {Klimo},\ and\ \citenamefont {Weber}}]{gelfer_prr24}%
  \BibitemOpen
  \bibfield  {author} {\bibinfo {author} {\bibfnamefont {E.}~\bibnamefont
  {Gelfer}}, \bibinfo {author} {\bibfnamefont {A.}~\bibnamefont {Fedotov}},
  \bibinfo {author} {\bibfnamefont {O.}~\bibnamefont {Klimo}}, \ and\ \bibinfo
  {author} {\bibfnamefont {S.}~\bibnamefont {Weber}},\ }\href@noop {}
  {\bibfield  {journal} {\bibinfo  {journal} {Physical Review Research}\
  }\textbf {\bibinfo {volume} {6}},\ \bibinfo {pages} {L032013} (\bibinfo
  {year} {2024}{\natexlab{b}})}\BibitemShut {NoStop}%
\bibitem [{\citenamefont {Niel}\ \emph
  {et~al.}(2018{\natexlab{a}})\citenamefont {Niel}, \citenamefont {Riconda},
  \citenamefont {Amiranoff}, \citenamefont {Lobet}, \citenamefont {Derouillat},
  \citenamefont {Pérez}, \citenamefont {Vinci},\ and\ \citenamefont
  {Grech}}]{Niel_2018}%
  \BibitemOpen
  \bibfield  {author} {\bibinfo {author} {\bibfnamefont {F.}~\bibnamefont
  {Niel}}, \bibinfo {author} {\bibfnamefont {C.}~\bibnamefont {Riconda}},
  \bibinfo {author} {\bibfnamefont {F.}~\bibnamefont {Amiranoff}}, \bibinfo
  {author} {\bibfnamefont {M.}~\bibnamefont {Lobet}}, \bibinfo {author}
  {\bibfnamefont {J.}~\bibnamefont {Derouillat}}, \bibinfo {author}
  {\bibfnamefont {F.}~\bibnamefont {Pérez}}, \bibinfo {author} {\bibfnamefont
  {T.}~\bibnamefont {Vinci}}, \ and\ \bibinfo {author} {\bibfnamefont
  {M.}~\bibnamefont {Grech}},\ }\href {\doibase 10.1088/1361-6587/aace22}
  {\bibfield  {journal} {\bibinfo  {journal} {Plasma Physics and Controlled
  Fusion}\ }\textbf {\bibinfo {volume} {60}},\ \bibinfo {pages} {094002}
  (\bibinfo {year} {2018}{\natexlab{a}})}\BibitemShut {NoStop}%
\bibitem [{\citenamefont {Niel}\ \emph
  {et~al.}(2018{\natexlab{b}})\citenamefont {Niel}, \citenamefont {Riconda},
  \citenamefont {Amiranoff}, \citenamefont {Duclous},\ and\ \citenamefont
  {Grech}}]{Niel_2018a}%
  \BibitemOpen
  \bibfield  {author} {\bibinfo {author} {\bibfnamefont {F.}~\bibnamefont
  {Niel}}, \bibinfo {author} {\bibfnamefont {C.}~\bibnamefont {Riconda}},
  \bibinfo {author} {\bibfnamefont {F.}~\bibnamefont {Amiranoff}}, \bibinfo
  {author} {\bibfnamefont {R.}~\bibnamefont {Duclous}}, \ and\ \bibinfo
  {author} {\bibfnamefont {M.}~\bibnamefont {Grech}},\ }\href {\doibase
  10.1103/PhysRevE.97.043209} {\bibfield  {journal} {\bibinfo  {journal}
  {Physical Review E}\ }\textbf {\bibinfo {volume} {97}},\ \bibinfo {pages}
  {043209} (\bibinfo {year} {2018}{\natexlab{b}})}\BibitemShut {NoStop}%
\bibitem [{\citenamefont {Jiang}\ \emph {et~al.}(2021)\citenamefont {Jiang},
  \citenamefont {Pukhov},\ and\ \citenamefont {Zhou}}]{pukhov_njp21}%
  \BibitemOpen
  \bibfield  {author} {\bibinfo {author} {\bibfnamefont {K.}~\bibnamefont
  {Jiang}}, \bibinfo {author} {\bibfnamefont {A.}~\bibnamefont {Pukhov}}, \
  and\ \bibinfo {author} {\bibfnamefont {C.~T.}\ \bibnamefont {Zhou}},\ }\href
  {\doibase 10.1088/1367-2630/ac0573} {\bibfield  {journal} {\bibinfo
  {journal} {New Journal of Physics}\ }\textbf {\bibinfo {volume} {23}},\
  \bibinfo {pages} {063054} (\bibinfo {year} {2021})}\BibitemShut {NoStop}%
\end{thebibliography}%

\end{document}